\documentstyle[psfig,referee]{l-aa-mpa}

\newcommand{\comment}[1]{}
        \def\smallskip{\vskip 2pt}

\begin{document}

\thesaurus{06(08.05.3; 08.16.4; 08.13.2) }

\title{Envelope burning over-luminosity:
a challenge to synthetic TP-AGB models} 

\author{ Paola Marigo$^{1,2}$ }
\institute{
$^1$ Max-Planck Institut f\"ur Astrophysik, Karl-Schwarzschild-Str. 1, D-87540 
      Garching bei M\"unchen, Germany \\
$^2$ Department of Astronomy, University of Padova,
      Vicolo dell'Osservatorio 5, 35122 Padova, Italy          
}


\date{Received May 1998 / Accepted .....}

\maketitle
\markboth{P.Marigo: Envelope burning over-luminosity: a challenge 
to synthetic TP-AGB models}{P.Marigo: Envelope burning over-luminosity: a challenge 
to synthetic TP-AGB models}


\begin{abstract}
Until recently synthetic AGB models had not taken into account the
break-down of the core mass-luminosity ($M_{\rm c} - L$) relation  
due to the occurrence of envelope burning in the most massive ($M
\ga 3.5 M_{\odot}$) and luminous ($M_{\rm bol} \la -6$) stars.

Marigo et al. (1998) made the first attempt to consistently include the
related over-luminosity effect (i.e. above the $M_{\rm c}-L$ 
relation) in synthetic TP-AGB
calculations. The method couples complete envelope integrations
with analytical prescriptions, these latter being presently
updated with the highly detailed relations 
by Wagenhuber \& Groenewegen (1998).
 
In this paper the reliability of the solution scheme 
is tested by comparison with the results of complete evolutionary
calculations for a $7 M_{\odot}$ AGB star undergoing envelope burning   
(e.g. Bl\"ocker \& Sch\"onberner 1991). 

Indeed,  
the method proves to be 
valid as it is able to reproduce with remarkable accuracy 
several evolutionary features  
of the $7 M_{\odot}$ star
(e.g. rate of brightening, luminosity evolution
as a function of the core mass and envelope mass for different
mass-loss prescriptions) 
as predicted by full AGB models.

Basing on the new solution method, we present extensive   
synthetic TP-AGB calculations for stars     
with initial masses of  
$3.5, \, 4.0, \, 4.5,$ and $5.0 \, M_{\odot}$, and three choices
of the initial metallicity, i.e. $Z=0.019$,
$Z=0.008$, and $Z=0.004$.
Three 
values of the mixing-length parameter are used, 
i.e. $\alpha=1.68, \, 2.0, \, 2.5$. 

We investigate the dependence of envelope burning
on such stellar parameters ($M$, $Z$, and $\alpha$).
The comparison between different cases 
gives hints on the interplay between envelope burning over-luminosity 
and mass loss, and related effects on TP-AGB lifetimes.


\end{abstract}  


\section{Introduction}
\label{intro} 
Synthetic AGB models may provide a powerful tool of investigation
(Renzini \& Voli 1981; Groenewegen \& de Jong 1993; Marigo et al. 1996a).
Evolutionary calculations can be easily carried out
over the whole mass range of AGB stars and for various metallicities,
 and the results  
tested according to the adoption of various
input prescriptions (e.g. analytical relations derived from full
AGB models,   
 mass-loss laws, parameters of the third dredge-up).
In this way,  it is possible 
to explore the sensitiveness of theoretical
predictions to different physical assumptions and, at the same time,   
to readily   
get an overall picture of aspects 
related both to single star
evolution 
(e.g. location on the H-R diagram, maximum AGB luminosities,
stellar lifetimes, 
initial mass-final mass relation, changes in the surface chemical
composition, stellar yields) 
and to integrated properties of the stellar aggregates in which AGB
stars are present 
(e.g. luminosity functions of oxygen- and carbon-rich AGB stars, 
contribution of evolved stars to the integrated light 
and to the chemical enrichment of the host galaxy).

Of course, the computational agility and flexibility typical of the synthetic
approach are paid with a certain loss of details if compared to
complete AGB  models. However, in order to get reliable results  
from synthetic analyses the degree of approximation in 
treating the physical processes must be
good enough so that no essential feature is missed.
 
Actually, a clear point of inadequacy of synthetic AGB models
has so far concerned the treatment of envelope burning in the most
massive stars  ($M \ga 3.5 M_{\odot}$).
In brief, 
the erroneous assumption common to AGB analyses 
performed either with the aid of convective envelope models
(Scalo et al. 1975; Renzini \& Voli 1981; Marigo et al. 1996a),
or purely analytical prescriptions (Groenewegen \& de Jong 1993)
is that the quiescent surface luminosity of a 
TP-AGB star experiencing envelope burning still  
obeys the standard  $M_{\rm c}-L$ relation, as 
in the case of 
lower mass stars ($M \la 3.5 M_{\odot}$).

On the contrary, over the years
complete AGB calculations have clearly indicated 
that nuclear
burning in the hottest convective envelope layers of   
the most massive TP-AGB stars may be a considerable 
energy source, making these stars depart 
significantly from the $M_{\rm c}-L$ relation towards higher
luminosities (Bl\"ocker \& Sch\"onberner 1991 (hereinafter also BS91);
Lattanzio 1992; Boothroyd \& Sackmann 1992; Vassiliadis \& Wood 1993;
Bl\"ocker 1995 (hereinafter also B95); Wagenhuber 1996).

Indeed, the recent discovery of the over-luminosity 
produced by envelope burning 
above the $M_{\rm c}-L$ relation 
 has notably
changed and improved our current understanding of the AGB phase,
setting  important implications.

In principle, the luminosity increase of a TP-AGB star 
is not bounded by
the classical AGB luminosity limit 
of $M_{\rm bol} \sim -7.1$, predicted by the Paczy\'nski (1970)
$M_{\rm c}-L$ relation 
for a mass of the degenerate core equal to the Chandrasekhar critical value 
of $\sim 1.4 M_{\odot}$ (BS91).
For this reason, the observation of AGB stars with 
luminosities close to $M_{\rm bol} \sim -7.1$ does not imply
that their core masses are close to $1.4 M_{\odot}$, a circumstance 
that has been until recently considered as a supporting indirect 
evidence
for the occurrence of Type I$_{1/2}$ Supernovae, consequent to
disruptive carbon ignition (see for example Wood et al. 1983).

Moreover, the  
 maximum quiescent luminosity, $L_{\rm max}$,
attained by an AGB star with envelope burning does not coincide
with the final luminosity at the so-called AGB tip, as in the case of
lower mass stars following the core mass-luminosity relation.
In higher mass stars the maximum of the luminosity occurs,
in practice, at the onset of the superwind, and hence before   
the AGB tip is reached (when the envelope has been almost completely
ejected and the core mass-luminosity relation is recovered).

It follows that at high luminosities, say $M_{\rm bol} < - 6.5$, the
core-mass luminosity relation cannot be longer employed in combination
with the initial mass-final mass  ($M_{\rm i}-M_{\rm f}$) relation to
estimate the age of a coeval system on the base of its brightest AGB
stars (usually inferred through the steps: $L_{\rm max} \rightarrow
M_{\rm f} \rightarrow M_{\rm i} \rightarrow$ age; see for instance the
review by Iben \& Renzini (1983)).

The over-luminosity effect is also expected to intensify the mass-loss 
process suffered by stars with envelope burning, so as to 
possibly anticipate
the onset of the superwind regime. Then, this would result in  
a reduction of the TP-AGB lifetimes and hence of the remnant
white dwarf masses. The latter consequence may concur to
solve the long-standing problem related to the
excess of white dwarfs more massive than $\sim 0.7 M_{\odot}$, as
predicted by synthetic AGB models (see Bragaglia et al. (1995) for
a punctual discussion of this point).
 
In this context, Marigo et al. (1998) first pointed out a
possible solution scheme to overcome the limit of synthetic models.
This paper aims at verifying the validity of the original 
treatment of envelope burning 
-- coupling 
the use of analytical relationships with complete
envelope integrations -- 
by testing its capability of reproducing
the results from full AGB calculations.

The general organisation of the paper is as follows.
In Sect.~\ref{method} the rationale underlying envelope
integrations is briefly recalled, together with the 
main analytical prescriptions.
In Sect.~\ref{comp} the validity of the method is checked by
comparison with full evolutionary calculations 
for a $7 M_{\odot}$ star performed by BS91 and B95.
Section~\ref{synt} presents the results of synthetic TP-AGB evolutionary calculations 
for stars with initial masses in the range 
$3.5 M_{\odot} \le M \le 5.0 M_{\odot}$, and initial metallicities 
$Z=0.019, \, Z=0.008,$ and $Z=0.004$.
The sensitiveness of envelope burning to 
stellar mass, metallicity, and mixing-length parameter is discussed in
Sect.~\ref{overlres}.
Some relevant quantities characterising the evolution of the model
stars are tabulated in Table~\ref{hbbtab}.  
Finally, 
Sect.~\ref{end} contains some concluding remarks and 
illustrates the intents 
of future works.

\section{Outline of the method}
\label{method}
The reader is referred to the work by Marigo et al. (1998)
for a detailed description of the method developed to calculate
the energy contribution from
envelope burning to the stellar luminosity.
Let us herein just summarise the basic points.

Given the total stellar mass $M$, the core mass $M_{\rm c}$, and the chemical composition
of the convective envelope at each time during the quiescent
inter-flash periods, 
the surface luminosity $L$ is singled out by means
of envelope integrations, provided that proper boundary conditions are
fulfilled. 

In this way, the erroneous assumption  $L = L_{\rm M_{\rm c}}$ -- 
where 
$L_{\rm M_{\rm c}}$ corresponds to the luminosity
predicted by the $M_{\rm c}-L$ relation for a given core mass -- is abandoned.
Moreover, 
the usual prescription, fixing $L_{r} = constant = L$ throughout 
the envelope, is
replaced with  the equation of energy balance under {\it static
approximation} (i.e. the entropy term
$- T \partial S/ \partial t$ is neglected):
\begin{equation}
\label{el}
\frac{\partial L_r}{\partial M_r}  =  \epsilon_r
\end{equation}
so that the complete
set of the stellar structure equations must be integrated.

In general, the determination of the unknown functions 
$r$, $P_{r}$, $T_{r}$, $L_{r}$ across the envelope  
 requires to specify  four boundary conditions.  
For any given pair of envelope parameters,
$L$ and $T_{\rm eff}$,
two conditions    
naturally derive from the integration of the photospheric 
equations for $T$ and $P$ down to the bottom of the photosphere
(Kippenhahn et al. 1967).
Then, since 
both $L$ and $T_{\rm eff}$ are actually free parameters in our
static envelope model,
two more boundary conditions must be fixed. 

To this aim, 
we express the quiescent surface luminosity $L$ as:
\begin{equation}
L  =  L_{\rm G} + L_{\rm He} - L_{\rm \nu} + L_{\rm H} + L_{\rm EB} 
\label{lumeq}
\end{equation} 
where $L_{\rm G}$ represents the rate of energy generation due to the
gravitational contraction of the core; $L_{\rm He}$ is the small
energy contribution from the He-burning shell; $L_{\rm \nu}$ is the
rate of energy loss via neutrinos; $L_{\rm H}$ and $L_{\rm EB}$ refer
to the rate of energy production by hydrogen burning in {\it
radiative} and {\it convective} conditions, respectively.

We remark that the gravitational contribution is 
included only in the energetic budget via the term $L_{\rm G}$,
but according to the {\it static approximation} used in our model, 
we do not actually take into account the possible contraction
and/or expansion of the structure during the evolution.

It is worth noticing that the energy contributions 
indicated in the right hand-side 
of Eq.~(\ref{lumeq}) are produced
within distinct regions of the star.
Specifically,  ($L_{\rm G} + L_{\rm He} - L_{\rm \nu}$) represents
the net rate of energy outflow from the core, this latter being commonly
 defined as the stellar interior below the H-He discontinuity.
Beyond the core,   
energy is produced by nuclear burning of hydrogen, at a total rate given by the
the sum of the two terms ($L_{\rm H}+L_{\rm EB}$).
It results that $L_{\rm EB} = 0$ in low-mass 
 TP-AGB stars ($M \la 3.5 M_{\odot}$)
complying with the $M_{\rm c}-L$ relation (i.e. without envelope burning).
Differently, the term $L_{\rm EB}$ can significantly contribute to the
energy budget of more massive TP-AGB stars 
($M > 3.5 M_{\odot}$), as the base
of the convective envelope penetrates into the H-burning shell.

In virtue of the site separation of the energy sources, 
it is convenient 
to set the two boundary conditions in question by specifying 
the local values of the luminosity $L_r$ at two
suitable transition points, 
provided that Eq.~(\ref{lumeq}) is satisfied. 
Denoting by $R_{\rm core}$ and
$R_{\rm conv}$ the radial coordinates of  the bottom of the H-burning shell
(i.e. where hydrogen abundance 
is zero, below which the mass $M_{\rm c}$ is contained)
and the base of the convective envelope, respectively,
we can write: 
\begin{eqnarray}
\label{l1}
L(r=R_{\rm core}) & = & L_{\rm G} + L_{\rm He} - L_{\rm \nu} \\
\label{l2}
L(r=R_{\rm conv}) & = &  L(r = R_{\rm core}) + L_{\rm H}
\end{eqnarray}
with the right hand-side members of both equations being known
functions of the core mass, envelope mass, and chemical composition
(see Sect.~\ref{analytic} for the adopted prescriptions).

Then, Eqs.~(\ref{l1}) and (\ref{l2}) together with the 
photospheric conditions provide the four boundary constraints
necessary to determine the entire envelope structure. 
Numerical integrations of the envelope  
are performed  with a very fine mass resolution,
the width of the innermost shells (where the structural
gradients become extremely steep)  typically amounting to 
$10^{-7} - 10^{-8} M_{\odot}$.

The  solution model yields, in particular,  
the quantity $L_{\rm EB}$, i.e. the energy produced within 
the convective envelope, and the     
($L$, $T_{\rm eff}$) pair, i.e. the current location 
on the H-R diagram,  without invoking further external assumptions.  

\subsection{Prescriptions for ($L_{\rm G}+L_{\rm He}-L_{\rm \nu}$) and
$L_{\rm H}$}
\label{analytic}
The boundary conditions expressed by Eqs.~(\ref{l1}) and (\ref{l2}) imply
the knowledge of the luminosity contributions:
\begin{itemize}
\item
($L_{\rm G} + L_{\rm He} - L_{\rm \nu}$) from the core
\item
$L_{\rm H}$ from radiative hydrogen burning 
\end{itemize}

To this aim, we adopt the analytical formulae
describing the light curves of TP-AGB stars, 
 presented by
Wagenhuber \& Groenewegen (1998). These prescriptions are
a high accuracy reproduction of the results  
from extensive grids of complete evolutionary calculations carried out by
Wagenhuber (1996) for stars with
initial masses in the range $0.8 M_{\odot} \le M \le 7.0 M_{\odot}$,
and metallicities $Z=0.0001$, $Z=0.008$, and $Z=0.02$. 
 
The maximum luminosity during quiescent H-burning is expressed 
as the sum of different terms:
\setcounter{equation}{0}
\renewcommand{\theequation}{5\alph{equation}}
\begin{eqnarray}
\label{lwag97_1}
L & = & (18160 + 3980 \log\textstyle{\frac{Z}{0.02}})(M_{\rm c} - 0.4468) \\
\label{lwag97_2}
  &   & + 10^{2.705+1.649 M_{\rm c}}   \\
\label{lwag97_3}
  &   & \times 10^{0.0237(\alpha-1.447) M_{\rm c,0}^{2} M_{\rm env}^{2} 
        (1-e^{-\Delta M_{\rm c}/0.01}) }  \\
\label{lwag97_4}
  &   & - 10^{3.529-(M_{\rm c,0}-0.4468) \Delta M_{\rm c}/0.01 }
\end{eqnarray}
\setcounter{equation}{5}
\renewcommand{\theequation}{\arabic{equation}}
where $Z$ denotes the metallicity, $M_{\rm c,0}$ is the core mass at
the first thermal pulse (see Table~\ref{hbbtab}), 
$\Delta M_{\rm c} = M_{\rm c} - M_{\rm c,0}$
gives the actual increment of the core mass, and $M_{\rm env}$ 
is the current envelope mass. Masses and luminosities are
expressed in solar units.

The first term (\ref{lwag97_1}) represents the usual linear $M_{\rm c}-L$ relation,
giving the quiescent luminosity of TP-AGB stars already in the 
full amplitude regime, with core masses
in the range $0.6 M_{\odot} \la M_{\rm c} \la 0.95 M_{\odot}$.
The second term (\ref{lwag97_2}) provides a correction becoming 
significant for high values of the core mass, $M_{\rm c} > 0.95 M_{\odot}$.
The third term (\ref{lwag97_3}) accounts for the over-luminosity produced
by envelope burning (i.e. the $L_{\rm EB}$ term of Eq.~(\ref{lumeq})), 
as a function of the envelope mass $M_{\rm env}$.
A dependence on the mixing-length parameter, $\alpha$, is included to reproduce 
the results from   
full calculations (i.e. Wagenhuber (1996) with $\alpha = 1.5$, 
Bl\"ocker (1995) with $\alpha = 2$, D'Antona \& Mazzitelli (1996)
with $\alpha = 2.75$). 
Finally, the fourth term (\ref{lwag97_4}) gives a negative correction to
the luminosity in order to mimic the sub-luminous and steep evolution
typical of the first pulses.

The nice distinction
between various terms in Eq.~(5) allows us to 
derive the term $L-L_{\rm EB}$, that is the 
the contribution of all energy sources but for envelope burning. 
This is an important point indeed,  
since $L_{\rm EB}$ is just the quantity
we aim at evaluating by means of envelope integrations as described in
Sect.~\ref{method}.
The luminosity produced by envelope burning can be eliminated from Eq.~(5),
by setting the corresponding term (\ref{lwag97_3}) equal to unity.
For the sake of clarity in notation, let us denote by $L_{\rm M_{\rm
c}}$ the resulting luminosity:
\setcounter{equation}{0}
\renewcommand{\theequation}{6\alph{equation}}
\begin{eqnarray}
\label{lmcrel1}
L_{\rm M_{\rm c}} & = & 
(18160 + 3980 \log\textstyle{\frac{Z}{0.02}})(M_{\rm c} - 0.4468) \\
\label{lmcrel2}
  &   &  + 10^{2.705+1.649 M_{\rm c}}   \\
\label{lmcrel3}
  &   & - 10^{3.529-(M_{\rm c,0}-0.4468) \Delta M_{\rm c}/0.01 } 
\end{eqnarray}
\setcounter{equation}{6}
\renewcommand{\theequation}{\arabic{equation}}
\noindent Hereinafter, we will refer to Eq.~(6) as the 
standard $M_{\rm c}-L$ relation adopted in this study.

This relation replaces the formulae by Boothroyd \&
Sackmann (1988a) and Iben \& Truran (1978), for
different ranges of the core mass, used in previous works (Marigo et
al. 1996ab, 1998).
The notable improvement is that the new prescription is based on 
homogeneous evolutionary calculations for a large range of core
masses and metallicities.
Moreover, it is worth remarking that
the quantity $L-L_{\rm EB}$ cannot be obtained     
when using the Iben \& Truran's formula (1978), which is
likely to
mask a non-quantifiable contribution from a weak envelope burning
(see Sect.~4.1 in Marigo et al. (1998)).

From the above scheme it follows that 
 the energy contribution from envelope 
burning, $L_{\rm EB}$, is supposed to satisfy  the relation: 
\begin{equation} 
\label{approxl} 
L = L_{\rm M_{\rm c}} + L_{\rm EB} 
\end{equation} 
In other words, we assume that the nuclear burning at the base of the convective
envelope produces the excess of luminosity above the underlying $M_{\rm c}-L$ relation.
 
According to  Wagenhuber (1996) it is possible to derive $L_{\rm H}$ from:
\begin{equation}
\label{frach}
\log\left(\frac{L_{\rm H}}{L}\right) = -0.012 - 10^{-1.25-113 \Delta M_{\rm c}}-0.0016 M_{\rm env}
\end{equation}
where the variables are
expressed in solar units. 
Since  the above relation is an analytical fit to
full calculations with no (or quite weak) envelope burning
(i.e. $L_{\rm EB} \sim 0$), we can 
safely assume $L=L_{\rm M_{\rm c}}$ in Eq.~(\ref{frach}).

Hence, we can write:
\begin{equation}
L_{\rm H} = f_{\rm H} L_{\rm M_{\rm c}}
\end{equation} 
where $f_{\rm H}$ is the fractional contribution of radiative H-burning
to $L_{\rm M_{\rm c}}$, and
\begin{equation}
\label{fracghe}
L_{\rm G}+L_{\rm He}- L_{\rm \nu} = (1-f_{\rm H}) L_{\rm M_{c}}
\end{equation}
is the complementary term
 to $L_{\rm M_{c}}$, including the gravitational
shrinking of the core, $L_{\rm G}$, the shell He-burning, $L_{\rm
 He}$, and neutrino losses, $L_{\rm \nu}$.

It turns out that, once the full amplitude regime has established, these
fractional luminosities attains typical values, slightly dependent
on the core mass, envelope mass, and metallicity, of 
$L_{\rm H}/L_{\rm M_{c}} \sim 0.96 \div 0.98$ and 
$(L_{\rm G}+L_{\rm He})/L_{\rm M_{c}} \sim 0.04 \div 0.02$.

Finally, it is worth recalling the general validity of the method,
which can be applied as well to solve the envelope structure of low-mass
stars without envelope burning (i.e. $L_{\rm EB} = 0$ and $L = L_{\rm
M_{\rm c}}$).

\subsection{Other analytical prescriptions}
\label{compdet}
The general structure of the synthetic TP-AGB model  
is the same as described in Marigo et al. (1996a, 1998).
With respect to these latter works, some input prescriptions have been updated
thanks to the recent re-determinations by Wagenhuber (1996) and
Wagenhuber \& Groenewegen (1998).
Besides the $M_{\rm c}-L$
relation already quoted in Sect.~\ref{analytic}, 
the other new analytical prescriptions are:

\begin{itemize}
\item The core mass - interpulse period relation: 
\end{itemize}
\setcounter{equation}{0}
\renewcommand{\theequation}{10\alph{equation}}
\begin{eqnarray}
\label{tipwag97_1}
\log t_{\rm ip} & = & (-3.628 + 0.1337 \:\: \textstyle\log\frac{Z}{0.02}) 
                      \:\: (M_{\rm c} - 1.9454) \\
\label{tipwag97_2}
                &   & - 10^{- 2.080 - 0.353 \:\: \log\frac{Z}{0.02} 
                      + 0.200 (M_{\rm env} + \alpha - 1.5)} \\
\label{tipwag97_3}
		&   & - 10^{ -0.626 - 70.30 \:\: (M_{\rm c,0} -
                      \log\frac{Z}{0.02})\:\: \Delta M_{\rm c} }
\end{eqnarray}
\setcounter{equation}{10}
\renewcommand{\theequation}{\arabic{equation}}
Here three components can be distinguished, namely:
the term~(\ref{tipwag97_1}) expresses the interpulse period
$t_{\rm ip}$ (in yr) as a decreasing exponential function of $M_{\rm c}$
with some dependence on the metallicity; the term~(\ref{tipwag97_2}) gives
a negative correction to include the effect of envelope burning
in somewhat reducing the inter-pulse period; and term
~(\ref{tipwag97_3}) reproduces the initial increase of $t_{\rm ip}$
starting from values that are shorter by almost a factor of two compared
to those derived from the asymptotic relation for the same core mass.    
\begin{itemize}
\item The rate of evolution of the hydrogen-exhausted core:  
\end{itemize}
\begin{equation}
\frac{d M_{\rm c}}{dt} = q \frac{L_{\rm H}}{X} \,\,\,\,\,\,\,\,\,
(M_{\odot}\,\, {\rm yr}^{-1})
\end{equation}
where
\begin{equation}
q = \left[(1.02 \pm 0.04)+0.017 \textstyle\frac{Z}{0.02}\right] 10^{-11}
\,\,\,\,\,\, ({\rm g~~erg}^{-1})
\end{equation}
with $X$ and $Z$ corresponding to the hydrogen and metal
abundances (mass fractions) in the envelope, respectively.
%

\begin{center}
\begin{figure} 
\centerline{
\psfig{file=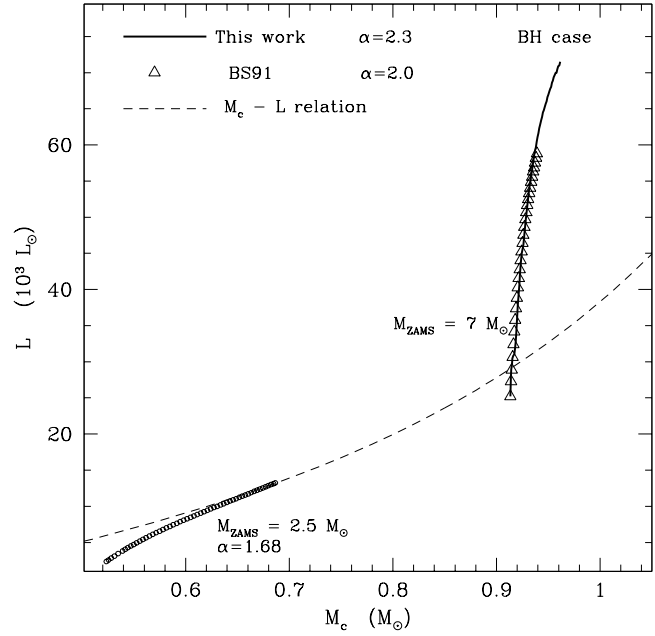,height=9.0truecm}} 
\caption{Evolution of the surface luminosity as a function of  
the core mass for a $7 M_{\odot}$ star with initial solar metallicity
(i.e. $Z=0.021$). 
The empty triangles 
refer to complete evolutionary calculations  
carried out by Bl\"ocker \& Sch\"onberner (1991) from the $1^{\rm st}$ 
up to the $30^{\rm th}$ thermal pulse.
The adopted mass-loss prescription is that suggested by Baud \& Habing
(1983; {\it BH case}).
For comparison, the luminosity evolution  predicted by our 
synthetic TP-AGB model is shown  
(solid line).  
Calculations are carried out  
starting from the same initial 
conditions at the first thermal pulse till the 
$60^{\rm th}$ pulse.
The dashed line corresponds to the underlying $M_{\rm c}-L$ relation 
adopted in this case, which is given 
by Eq.~(6) multiplied 
by a factor of $1.16$.
The evolution of a $2.5 M_{\odot}$ star with no  
envelope burning is also plotted. 
See the text for more details.} 
\label{block70_bh} 
\end{figure} 
\end{center}

\section{Comparison with full calculations}
\label{comp}
%
\begin{center}
\begin{figure}
\centerline{ 
\psfig{file=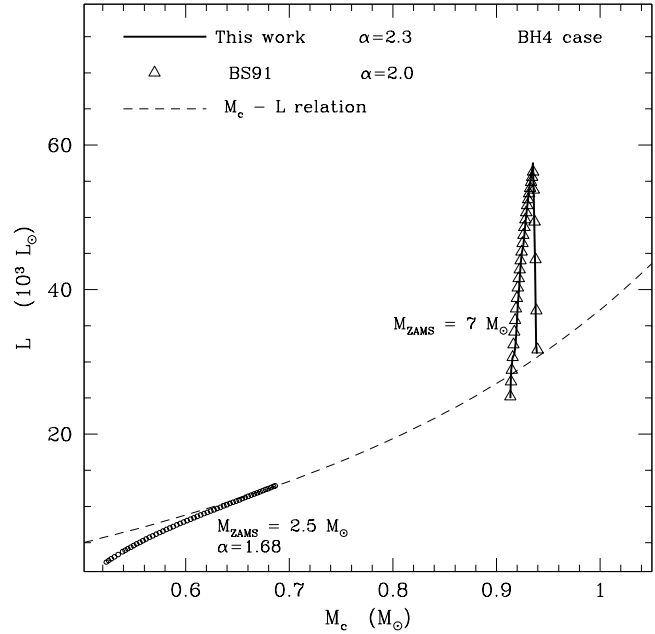,height=9.0truecm}} 
\caption{The same as in Fig.~\protect\ref{block70_bh4}, but with a
different prescription for mass-loss ({\it BH4 case}).
The Baud \& Habing's formula (1983), applied since the beginning  of the
AGB, is then replaced with a constant rate, $\dot M = 4 \times 10^{-4}
 M_{\odot}$yr$^{-1}$,
as soon the core mass has grown up to $0.93537 M_{\odot}$.} 
\label{block70_bh4} 
\end{figure} 
\end{center}
%
The most meaningful test to check
the reliability of the method developed to account envelope burning
is to compare the predictions of the synthetic TP-AGB models with  
those from complete AGB evolutionary calculations.
We consider the results   
obtained
by BS91 and B95  
for a TP-AGB star with initial mass of $7 M_{\odot}$ and chemical 
composition $[X=0.739, Y=0.240, Z=0.021]$.

We calculate the synthetic TP-AGB evolution of the $7 M_{\odot}$ star 
starting from the
same initial conditions at the first thermal pulse as in BS91, i.e. with  
$M = 6.871 M_{\odot}$, $M_{\rm c} = 0.91335 M_{\odot}$, $L = 25217 L_{\odot}$. 
Specifically, the underlying  $M_{\rm c}-L$ relation is  
that given by Wagenhuber \& Groenewegen (1998) (Eq.~(6)), multiplied by a proper  
factor of $1.16$. This latter 
is  calibrated in order to obtain the same 
value of the luminosity at the first thermal pulse as in BS91.

In our envelope model the mixing-length parameter is set equal to 
$\alpha = 2.3$ so as to obtain values of the effective 
temperature similar to those derived from full calculations (see
Fig.~5 in B95), 
which were   
actually carried out with a lower value, $\alpha =2.0$.
This is most likely due to the different opacities employed in
Bl\"ocker's calculations
(Cox \& Stewart 1970), and in the present study (Iglesias \& Rogers 1996;
Alexander \& Ferguson 1994).

As far as mass loss by stellar winds on the AGB is concerned, two are
the prescriptions here adopted as in B95, so that we distinguish two cases:
\begin{enumerate} 
\item 
the {\it BH case} 
referring to the use of the Baud \& Habing's (1983) 
modification of Reimers'
formula; 
\item  
the {\it BH4 case} 
corresponding to the use of the Baud \& Habing's law (1983) from
the beginning of the AGB  
till a certain stage (i.e. when $M_{\rm c}=0.93537 M_{\odot}$ at
the $26^{\rm th}$ pulse), 
beyond which a constant rate of $4 \times 10^{-4} M_{\odot}$ yr$^{-1}$ 
is artificially introduced to mimic the onset of the superwind
regime. 
\end{enumerate}
For more details the reader should refer to B95.
\begin{table*} 
\centering 
\caption{A solar-metallicity $7 M_{\odot}$ star with envelope burning: 
comparison between full TP-AGB modelling performed 
by Bl\"ocker \& Sch\"onberner (1991) (BS91; {\it BH case}) and our calculations (M98), 
for four selected values of the core mass.} 
\label{block70tab} 
\begin{tabular}{ccclcccccc} 
\hline 
\noalign{\smallskip} 
\multicolumn{1}{c}{BS91/M98} & 
\multicolumn{1}{c}{$\frac{M_{\rm c}}{M_{\odot}}$} & 
\multicolumn{1}{c}{$\frac{M}{M_{\odot}}$} & 
\multicolumn{1}{c}{$\frac{\dot M}{M_{\odot}{\rm yr}}$} & 
\multicolumn{1}{c}{$\frac{L}{L_{\odot}}$} & 
\multicolumn{1}{c}{$|\epsilon_{L}|$} & 
\multicolumn{1}{c}{$\frac{L_{\rm EB}}{L_{\rm H} + L_{\rm EB}}$} & 
\multicolumn{1}{c}{$\frac{\dot M_{\rm bol}}{{\rm mag \, yr}}$} \\ 
\noalign{\smallskip} 
\hline 
\noalign{\medskip} 
BS91 &  0.91335 &  6.871  &  4.9$\:10^{-7}$  &  25217  &       &  0.02  &  -3.3$\: 10^{-5}$  \\ 
M98  &  0.91335 &  6.871  &  6.8$\:10^{-7}$  &  25217  & 0.000 &  0.01  &  -5.5$\: 10^{-5}$  \\ 
BS91 &  0.92061 &  6.854  &  1.1$\:10^{-6}$  &  40268  &       &  0.30  &  -1.4$\: 10^{-5}$  \\ 
M98  &  0.92061 &  6.850  &  1.1$\:10^{-6}$  &  37788  & 0.062 &  0.33  &  -1.5$\: 10^{-5}$  \\ 
BS91 &  0.92970 &  6.821  &  1.7$\:10^{-6}$  &  51662  &       &  0.46  &  -7.2$\: 10^{-6}$  \\ 
M98  &  0.92970 &  6.807  &  1.9$\:10^{-6}$  &  51537  & 0.003 &  0.46  &  -7.6$\: 10^{-6}$ \\    
BS91 &  0.93909 &  6.774  &  2.1$\:10^{-6}$  &  58806  &       &  0.52  &  -4.0$\: 10^{-6}$  \\ 
M98  &  0.93909 &  6.745  &  2.6$\:10^{-6}$  &  60760  & 0.033 &  0.51  &  -3.4$\: 10^{-6}$  \\ 
\noalign{\smallskip} 
\hline 
\noalign{\medskip} 
\end{tabular} 
\small 
\begin{itemize} 
\item  $M_{\rm c}$:  hydrogen-exhausted core 
\item  $M$: current stellar mass 
\item  $\dot M$: mass-loss rate 
\item  $L$: quiescent surface luminosity 
\item $| \epsilon_{L} | =\textstyle{|\frac{L_{\rm M98}-L_{\rm
BS91}}{L_{\rm BS91}}|}$: percentage difference in the evaluation  
of $L$ (at given $M_{\rm c}$) 
derived in the present work with respect to full calculations 
\item  $\frac{L_{\rm EB}}{L_{\rm H}+L_{\rm EB}}$: fraction of the  
total hydrogen luminosity 
provided by envelope burning 
\item $\dot M_{\rm bol}$: rate of brightening
\end{itemize} 
\end{table*}

Figures~\ref{block70_bh} 
(for the {\it BH case}) and \ref{block70_bh4} (for the {\it BH4 case})
show that the results of full calculations are remarkably well reproduced 
by our synthetic calculations. The displayed luminosity evolution 
refers to the pre-flash maximum values before the occurrence of each
thermal pulse. 
The $7 M_{\odot}$ star starts to  
depart from the $M_{\rm c}-L$ as soon as it enters the TP-AGB phase, 
quickly increasing its luminosity because of 
the occurrence of envelope burning.
 
In the {\it BH case} the luminosity is still steeply increasing
when calculations have been interrupted (at the $30^{\rm th}$ pulse in BS91;
at the $60^{\rm th}$ pulse in this work), due to the fact that the
stellar mass has not yet been significantly reduced by stellar winds.
Differently, 
the artificial onset of a superwind mass-loss rate in the {\it BH4 case} 
causes the quick ejection of the envelope over  the subsequent $4-5$
interpulse periods.
Our synthetic calculations quite well reproduce the overall features 
of the luminosity evolution, 
comprising the
initial rising, the luminosity peak, the decline
following the activation of the high mass-loss rate, and the final
re-approaching towards the $M_{\rm c}-L$ relation.

For purpose of comparison, 
the predicted TP-AGB evolution of a 
solar-metallicity $2.5 M_{\odot}$ star is also plotted. 
No evidence of departure from 
the standard $M_{\rm c}-L$ relation shows up in this case. 

As far as the $7 M_{\odot}$ star is concerned,
different  
properties can be then compared, namely: i) the surface luminosity for given core mass 
and envelope mass; 
ii) the rate of brightening $\dot M_{\rm bol}$;  
iii) the fraction of  
the hydrogen luminosity produced at the hot base of the 
convective envelope $L_{\rm EB}/(L_{\rm H} + L_{\rm EB})$;
iv) the current total mass; and v) the mass-loss rate.   
These quantities are indicated in Table~\ref{block70tab} at fixed  
values of the core mass for four selected thermal pulses 
 ({\it BH case}; in analogy with Table~1 of BS1).

\begin{center}
\begin{figure}
\centerline{
\psfig{file=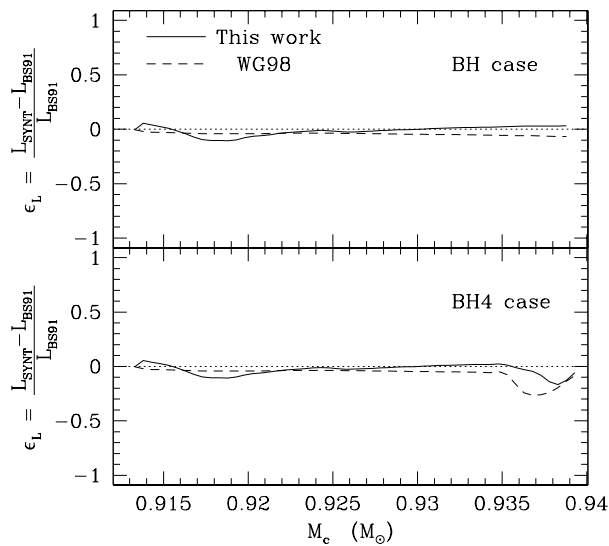,height=9truecm}}
\caption{Percentage difference in the estimation of the quiescent
surface luminosity according to synthetic AGB calculations ($L_{\rm
SYNT}$) (by envelope integrations in this work (M98, solid line), and
with the analytical fit of Eq.~(5) by Wagenhuber \& Groenewegen (1998)(WG98,
dashed line)), with respect to the results from complete calculations
($L_{\rm BS91}$), as a function of the core mass for an evolving $7
M_{\odot}$ star.  Top and bottom panels refer to the {\it BH case} and
{\it BH4 case} for mass loss, respectively.}
\label{diffl}
\end{figure}
\end{center}
 
The agreement is indeed satisfactory. 
The relative errors,  
$|\epsilon_{L}| = |L_{\rm M98}-L_{\rm BS91}|/L_{\rm BS91}$, 
of the surface luminosity estimated with our model, $L_{\rm M98}$,  
with respect to 
that obtained in the reference work, $L_{\rm BS91}$,
 do not exceed few percents in all cases.
The very nice accordance can be better appreciated from 
Fig.~\ref{diffl}, showing $\epsilon_{L}$ 
(solid line) as a function of the core
mass for the entire sequence of thermal pulses calculated  by
BS91 for the {\it BH case} (top panel), and B95 for the {\it BH4 case}
(bottom panel).

The values of $\epsilon_{L}$ are always quite small, also in consideration 
that the models  are intrinsically 
different just because of the use of 
different input physics (e.g. opacities).
Moreover, it is worth noticing that in the {\it BH case} the greatest differences in
luminosity for given core mass occur just in correspondence to the first thermal pulses,
the behaviour of which is usually irregular and model dependent, 
and then hardly reproducible.
In the {\rm BH4 case} $\epsilon_{\rm L}$ is mostly confined to few
percents as well.
An increasing trend (to modest values
still) shows up in the very late few interpulse periods, which are
characterised by the drastic
reduction of the envelope mass. 
The reason of this may be ascribed to   
somewhat different core mass-interpulse period relations (in
Wagenhuber's (1996) and BS91 models). Then, given the extremely high
mass-loss rate in these stages -- when envelope burning 
is powering down but still operating -- even a small a difference in the
core mass just prior the occurrence of a thermal pulse can correspond
to rather different values of the envelope mass, and hence of the
expected surface
luminosity.

A similar check is performed using  the 
analytical fit suggested by Wagenhuber \& Groenewegen (1998; WG98) and
quoted in Eq.~(5). The term (\ref{lwag97_3}), expressing the luminosity
contribution from envelope burning, is evaluated by setting $\alpha =
2.0$, as employed by BS91.    
The agreement with full calculations also turns out to be really good in
both cases as illustrated in Fig.~\ref{diffl}.
   
However, the advantage of our method, based on envelope integrations 
during the evolutionary calculations, is that it not only gives an  
estimate of $L_{\rm EB}$ and hence of the actual stellar luminosity
(as the analytical fit by WG98 does), but
the entire envelope structure of the star is consistently determined
at each time step. This aspect is relevant, for instance, to the
analysis of the nucleosynthesis due to envelope burning, which is followed in
detail by integrating the nuclear network,
once the density and
temperature stratifications across the envelope are known.
Moreover, the envelope model allows to check the sensitivity of the
results to possible changes of the input physics (e.g. opacities,
nuclear rates, mixing scheme).
 
\begin{center}
\begin{figure}
\centerline{
\psfig{file=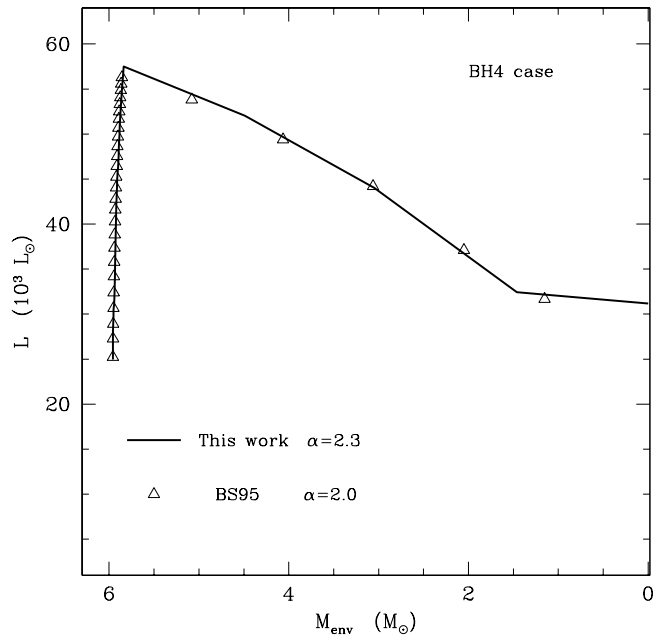,width=9.0truecm}}
\caption{Evolution of the surface luminosity  for the $7 M_{\odot}$ AGB model
as a function of the envelope mass, $M_{\rm env}$, corresponding
to {\it BH4 case} for mass loss.
The notation is the same as in Fig.~\protect\ref{block70_bh}.
Note the quick decline of the surface luminosity after the onset of
the superwind.}
\label{env}
\end{figure}
\end{center}

Finally, Fig.~\ref{env} shows the surface luminosity of
the $7 M_{\odot}$ star as a function of the envelope mass,
with reference to the {\it BH4 case} for mass loss (in analogy with Fig.~9 in B95).
The agreement of synthetic results (solid line) with complete calculations 
(empty triangles) is again very good, indicating that the method 
is reliable to estimate the current strength of  
envelope burning as the envelope mass is being reduced by stellar winds.    

\section{Synthetic evolutionary calculations}
\label{synt}
The present paper is the first of a series devoted to present the
results coming out from an extensive synthetic analysis 
of the TP-AGB evolution.  

Evolutionary calculations have been carried out to
follow the TP-AGB phase 
(from the first thermal pulse until the ejection of the envelope)
for a dense grid stellar models with initial
masses in the range $0.8 M_{\odot} \div 5 M_{\odot}$, and  
three values  of the original
chemical composition, $[X=0.708, Y=0.273, Z=0.019]$, $[X=0.742, Y=0.250,
Z=0.008]$, and $[X=0.756, Y=0.240, Z=0.004]$ (Marigo 1998, PhD Thesis).
For each stellar model the initial conditions at the first thermal
pulse are extracted from full calculations 
performed by means of the Padua stellar evolution code 
(Girardi \& Bertelli 1998; Girardi et al. 1998, in preparation).
We remind the reader that the inclusion
of moderate overshoot from core and external convection (Chiosi et al. 1992;
Alongi et al. 1993) leads to a lower value of the maximum mass
($M_{\rm up} \sim 5 M_{\odot}$)
for a star to pass through the AGB phase, than predicted by  classical
models without overshooting ($M_{\rm up} \sim 7-8 M_{\odot}$).

The analytical ingredients of the TP-AGB model are the same ones
as in Marigo et al. (1996a, 1998), with some updates already 
presented in Sects.~\ref{analytic} and \ref{compdet}.
We recall that the adopted prescription for mass loss is that
by Vassiliadis \& Wood (1993).  
The basic physical inputs employed in the static envelope model are 
the following.
Nuclear reaction rates are those 
compiled by Caughlan \& Fowler (1988), the screening factors are given
by Graboske et al. (1973). At high temperatures ($T > 10^{4} K$) stellar opacities
are taken from Iglesias \& Rogers (1996) (OPAL), at low temperatures ($T< 10^{4}$)
the opacity tables by Alexander \& Ferguson (1994) are used.\
The value  $\alpha=1.68$ adopted in envelope integrations derives from
the calibration of the solar model (Girardi et al. 1996), which we
usually refer to as the {\it standard case}. 
For purpose of comparison, other values are also used   
in the computation of envelope models (i.e. $\alpha=2.0$ and $\alpha=2.5$).

\subsection{An improved treatment of the third dredge-up}
\label{dred3}
The usual 
treatment of the third dredge-up in synthetic calculations 
is based on the adoption of two free
parameters, namely:  the efficiency $\lambda$, and the minimum core
mass for
convective dredge-up $M_{\rm c}^{\rm min}$. 
They are calibrated so as to fit some basic observational constraint,
e.g. the observed luminosity function of carbon stars
in the LMC ($M_{\rm c}^{\rm min} = 0.58 M_{\odot}$ 
and $\lambda \sim 0.6-0.7$ according to Groenewegen \& de Jong (1993) 
and Marigo et al. (1996a)).
The main purpose is to provide useful indications on 
the real occurrence of convective dredge-up, 
given the large degree of uncertainty still affecting the present 
understanding of this process in complete analyses of thermal pulses.

However, a weak point of synthetic models is the
assumption that both dredge-up parameters are constant, regardless
of the stellar mass and metallicity. 
On the contrary, complete models of AGB stars indicate that 
 the onset of dredge-up 
(related to $M_{\rm c}^{\rm min}$) and its efficiency
(related to $\lambda$)
is favoured in stars of higher masses and lower metallicities
(Wood 1981; Boothroyd \& Sackmann 1988b).

In this work, this limitation of synthetic models is partially
overcome. 
In Marigo et al. (1998, in preparation) an exhaustive description of
the method is presented. Suffice it to formulate here the basic concepts.

According to the results from detailed
calculations of thermal pulses
it turns out that, at the stage of
the post-flash luminosity peak,
the penetration of envelope convection into the inter-shell region would
occur only if the base temperature, $T_{\rm b}$, 
approaches or exceeds
some critical value $T_{\rm b}^{\rm dred}$, 
which turns out to be almost independent from $M_{\rm c}$ and $Z$ 
(e.g. $\log T_{\rm b}^{\rm dred} \sim 6.7$
as indicated by Wood (1981); $\log T_{\rm b}^{\rm dred} \sim 6.5$
according to Boothroyd \& Sackmann 1988b).
As already pointed out by Wood (1981),
this useful  indication provides the 
basic criterion to infer {\it if and when}
dredge-up takes place.
The use of $M_{\rm c}^{\rm min}$ is then abandoned.
Of course, the critical value for $T_{\rm b}^{\rm dred}$ is a free
parameter as well, which must be calibrated on the basis of
some observational constraint.

However, the improvement is real.
Every time a thermal pulse is expected during calculations, envelope
integrations are performed to check whether the condition on the base
temperature is satisfied. It follows that, contrary to the test 
based on the constant $M_{\rm c}^{\rm min}$ parameter, with this scheme
the response depends not only on the core mass but also on the current physical
conditions of the envelope (i.e. the surface luminosity peak, the effective
temperature, the mass, the chemical composition).
Moreover, this method allows to determine not only  
the onset,  but also the possible shut-down of dredge-up occurring 
when the envelope mass is significantly reduced by stellar winds.

As already mentioned, the dredge-up parameters, 
$\lambda$ and $T_{\rm b}^{\rm dred}$, need to be specified.
The carbon star luminosity functions (CSLFs) in the LMC and SMC 
provide the 
observational constraint.
According to Marigo (1998, Phd Thesis),
the calibration for the LMC  
yields $\lambda=0.50$ and $T_{\rm b}^{\rm dred}=6.4$,
whereas for the SMC we get $\lambda=0.65$ and $T_{\rm b}^{\rm
dred}=6.4$.
Then, a higher efficiency of convective dredge-up seems
to be required at lower metallicities, which would agree with
predictions by full calculations. 

In this way we fix the values of the dredge-up
parameters adopted during the present calculations for different
metallicity sets. Specifically, the calibration based on the CSLF in
the LMC is applied to TP-AGB models with metallicities $Z=0.008$, 
the reproduction of the CSLF in the SMC specifies the 
 dredge-up parameters for TP-AGB models with $Z=0.004$.
The former calibration is applied also to the metallicity set $Z=0.019$,
even if a lower $\lambda$ would be suggested by extrapolating
from the $Z=0.008$ and $Z=0.004$ cases.

Regarding the stellar models studied in this work
($3.5 M_{\odot} \le M \le 5.0 M_{\odot}$), 
it turns out that
envelope burning, as expected, prevents the conversion to carbon stars
for most of their TP-AGB phase because of the efficient nuclear
transmutation of newly dredged-up carbon into nitrogen.  Possible
transitions to the C-class may occur either early in the evolution
when dredge-up still dominates over weak (but growing in efficiency) 
envelope burning, or
in the very late stages as envelope burning extinguishes and a few
more dredge-up events possibly take place.  All these aspects are
analysed and discussed in Marigo et al. (1998, in preparation).

\subsection{Evolution in the $M_{\rm c} - L$ diagram}
\label{lmcres}
\begin{center}
\begin{figure}
\centerline{
\psfig{file=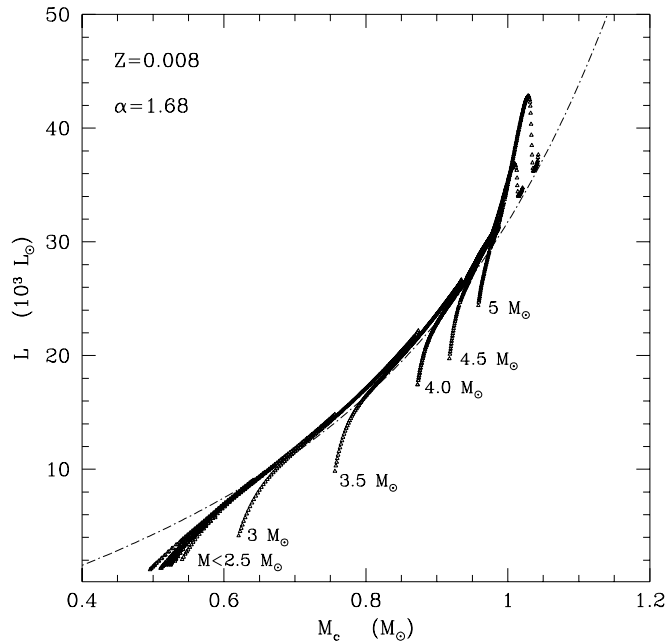,width=9truecm}}
\caption{Quiescent luminosity evolution as a function
of the core mass for TP-AGB stars with initial
chemical composition $[X=0.742, Y=0.250, Z=0.008]$, and
mass as indicated nearby the corresponding track.
Symbols correspond to the pre-flash
luminosity maximum before the occurrence of each thermal pulse.
The dot-dashed line represents the $M_{\rm c}-L$ relation 
for $Z=0.008$.}
\label{lmcz008}
\end{figure}
\end{center}

Figure~\ref{lmcz008} shows the luminosity evolution described
by TP-AGB stars with initial metallicity $Z=0.008$ as a function of 
core mass, according to the {\it standard case} (i.e. $\alpha =1.68$). 
  Each symbol along the curves refers
to the quiescent pre-flash luminosity maximum, just before
the occurrence of a thermal pulse.
The dot-dashed line corresponds to the standard $M_{\rm c}-L$ relation
valid for  full amplitude regime with no envelope burning.
It is  
expressed by Eq.~(6) with the term (\ref{lmcrel3}) set equal to zero, as this
refers to the first pulses which are not yet in the asymptotic regime.   

Three points are worthy to be remarked.
First, all the tracks are characterised by an initial sub-luminous
evolution typical of the first thermal pulses, below the
$M_{\rm c}-L$ relation.
Second, as soon as the standard relation is approached, 
stars with mass $M \la 3.5 M_{\odot}$ increase their luminosity 
closely following    
the same relation till
the end of evolution, 
regardless of the current value of the envelope mass.
Quite a different behaviour characterises 
the luminosity evolution of more massive stars ($3.5 M_{\odot} < M \le 5 M_{\odot}$),
departing from the $M_{\rm c}-L$ relation because of the 
occurrence of envelope burning. This aspect will be 
 discussed in detail
in Sect.~\ref{overlres}.
Third, we can  notice that 
in the cases with no envelope burning, 
stars may not exactly obey the relation shown
in Fig.~\ref{lmcz008}, but 
evolve towards slightly
higher luminosities, in particular during the last evolutionary
stages.

\begin{center}
\begin{figure}
\centerline{
\psfig{file=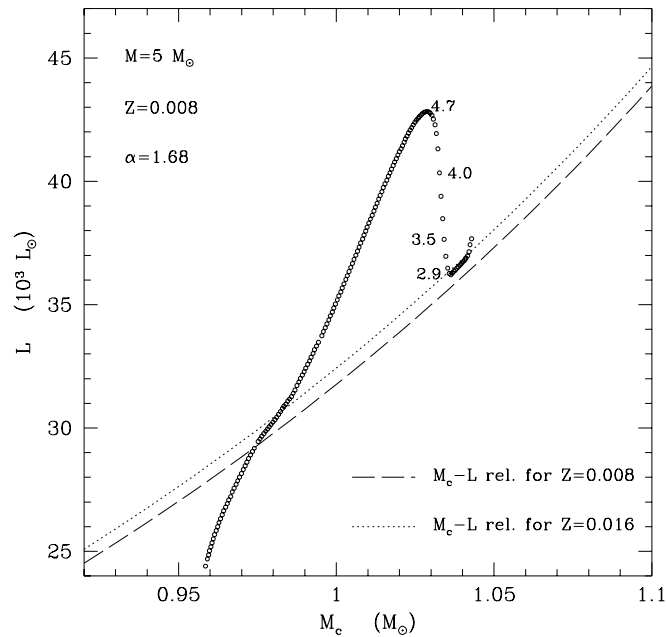,height=9.0truecm}}
\caption{Quiescent luminosity evolution of a $5.0 M_{\odot}, Z=0.008$ star
experiencing envelope burning. The open circles correspond
to the maximum quiescent luminosity before each He-shell flash.
The numbers along the 
curve indicate the current stellar mass in solar units.
The dashed and dotted lines represent the reference $M_{\rm c}-L$ relation
for $Z=0.008$  and $Z=0.016$, respectively.}
\label{50z008eb}
\end{figure}
\end{center}

This can be explained considering that the plotted $M_{\rm c}-L$ 
relation refers to a given value of the metallicity, $Z=0.008$,
so that the weighting factor, $3980 \log(Z/0.02)$, in the term
(\ref{lmcrel1}) is constant.
However, during calculations  
we take into account the possible increase of
the effective metallicity, defined as $Z=1-X-Y$,
due to
the surface chemical enrichment produced by convective dredge-up,
and possibly by envelope burning.
Therefore, for a given core mass the luminosity is 
expected to be higher at increasing metallicity. The effect is
more pronounced after the last dredge-up episodes, 
when the dilution of newly synthesised 
elements involves a smaller residual envelope mass.
To this respect, an example is illustrated in Fig.~\ref{50z008eb},
referring to the $(5 M_{\odot}; Z=0.008)$ TP-AGB model experiencing
both dredge-up events and envelope burning (see also Sect.~\ref{ebm}). 
Similar results hold for the other two sets of TP-AGB models 
here considered (with  metallicities $Z=0.019$ and $Z=0.004$).

\section{Over-luminosity produced by envelope burning}
\label{overlres}
%
%
%

As already mentioned, the energy contribution from envelope burning makes
massive TP-AGB stars leave the $M_{\rm c}-L$ relation during part of their
evolution. 
In this study we analyse the dependence of envelope burning
on three basic parameters namely, the mass of the star $M$, 
the metallicity $Z$ of the envelope,  
and the mixing-length parameter, $\alpha=\Lambda/H_{P}$,
where $\Lambda$ is the mixing length according to the classical theory of 
convection (Mixing Length Theory; B\"ohm-Vitense 1958), and
$H_{P}$ is the pressure scale-height.
The results are clearly illustrated in 
Figs.~\ref{extralz02} -- \ref{extralz004}, referring to TP-AGB stars 
with initial mass of $3.5, \, 4.0, \,4.5,$ and $5.0 M_{\odot}$ for
three values of the initial metallicity, $Z=0.019, \,  Z=0.008,$ and
$Z=0.004$, respectively.
For each star of given mass and chemical composition,
envelope integrations are carried out adopting three values of
the mixing-length parameter, 
$\alpha = 1.68$ ({\it standard case}),  $\alpha=2.00$, and
$\alpha=2.50$.

We can notice that for particular combinations of the stellar
parameters here considered (i.e. lower $M$, higher $Z$, and lower
$\alpha$, as in the ($3.5 M_{\odot}, \, Z=0.019, \, \alpha=1.68$)
model of Fig.~\ref{extralz02}), envelope burning does not develop or
it is so weak that the over-luminosity effect does not show up and the
the evolution of the quiescent luminosity complies with the $M_{\rm
c}-L$ relation.
Moreover, we specify that the results for the $(5.0 M_{\odot}, Z=0.004)$ model 
are not actually present since this
is just a limit case. The stellar mass slightly exceeds the critical
value $M_{\rm up}$ for that metallicity, so that carbon
ignition has occurred before the onset of the thermally  pulsing regime.

It is worth remarking that
$M$, $Z$, and $\alpha$ 
significantly  affect the luminosity evolution of 
a star with envelope burning as long as the stellar mass
is not so dramatically lowered, before the  
re-approaching onto the $M_{\rm c}-L$ relation. 
This corresponds to most 
of the TP-AGB duration, since 
high mass-loss rates are typically attained
at the very end of the evolution (Vassiliadis \& Wood 1993).
Once the super-wind regime has developed,
the subsequent luminosity evolution is crucially controlled  
by the efficiency of mass loss in reducing the envelope
mass down to the complete ejection.
An earlier onset of the super-wind would favour an earlier extinction
of envelope burning and,  correspondingly, a lower maximum luminosity
would be reached by the star before the subsequent decline towards 
the $M_{\rm c}-L$ relation.
In principle,  the onset of the super-wind regime
as soon as the star enters the AGB phase may even prevent
the occurrence of envelope burning.

In the following discussion we will consider the effects
produced by varying $M$, $Z$, and $\alpha$, for a given mass-loss
prescription  (i.e. Vassiliadis \& Wood 1993). 
The main results relevant to the present analysis are
indicated in Table~\ref{hbbtab}.
\subsection {The dependence on $M$} 
\label{ebm}
For fixed $Z$ and $\alpha$, envelope burning is expected
to be more efficient in stars of greater mass.
From Figs.~\ref{extralz02} -- \ref{extralz004} 
it turns out that
the maximum excursion  of
the luminosity  above  the $M_{\rm c}-L$ relation is higher in
initially more massive models
(see also the $M_{\rm bol}^{\rm peak}$ enter in Table~\ref{hbbtab}).
Moreover, the development of envelope burning crucially
depends on the current envelope mass during the evolution.
This effect is exemplified in Fig.~\ref{50z008eb}, referring to
the case of the ($5 M_{\odot}, Z=0.008, \alpha=1.68$) model.
This behaviour of the luminosity agrees
with the results from full evolutionary calculations of the TP-AGB phase
(e.g. B95; Boothroyd \& Sackmann 1992; Vassiliadis \& Wood 1993).

Envelope burning develops since the first sub-luminous interpulse
periods, so that when thermal pulses attain the full-amplitude regime
the star does not settle on the $M_{\rm c}-L$ relation (dashed line),
but quickly reaches higher and higher luminosities, with a brightening
rate (few times $-10^{-6} \div -10^{-5}$ mag yr$^{-1}$; see the $\dot
M_{\rm bol}$ entry in Table~\ref{hbbtab}), much greater than expected
from the slope of the standard relation (typically $-7 \div -9 \times
10^{-7}$ mag yr$^{-1}$).

The luminosity growth goes on as long as 
the envelope remains massive enough to support
high envelope base temperatures ($T_{\rm b} > 40-60 \times 10^{6} \, K$).
In fact, the luminosity decline after
the maximum is concomitant with the onset of the super-wind, 
when the star starts to rapidly lose mass at significant rates
(i.e. $\dot M \sim 10^{-5} \div  10^{-4} M_{\odot}$~yr$^{-1}$).
This drastically weakens the efficiency of envelope burning so that, 
finally, the over-luminosity vanishes 
and the star approaches again the $M_{\rm c}-L$ relation where
it remains during the very last stages till the end of evolution.
To this respect, we can notice that the final recovering of the
$M_{\rm c}-L$ relation (dotted line) 
is consistent with an effective  metallicity 
($Z=0.016$) that
is twice the value at the beginning of the TP-AGB phase
($Z=0.008$; dashed line), because of the 
occurrence of both dredge-up and envelope burning.

In Table~\ref{hbbtab} we also indicate the value of the envelope mass 
at the shut-down of envelope burning ($M_{\rm env}({\rm no EB})$ entry),
i.e. when its contribution to
the stellar luminosity has decreased to less than $1$~\%.
Note that $M_{\rm env}({\rm no EB})$ is  
not constant at all, but it varies mostly in the range between roughly $2 M_{\odot}$
 and $0.5 M_{\odot}$. It results that the more efficient 
envelope burning has been (e.g. for larger $M$, lower $Z$, higher $\alpha$),
the smaller is this critical value of the envelope mass.
  
%
\begin{table*}
\caption{Properties of TP-AGB stars with envelope burning.}
\label{hbbtab}
\centering
\begin{tabular}{cccccccccc}
\hline
\noalign{\smallskip}
\multicolumn{1}{c}{$Z$} &
\multicolumn{1}{c}{$\alpha$} &
\multicolumn{1}{c}{$M_{\rm i}$} &
\multicolumn{1}{c}{$M_{\rm c,0}$} &
\multicolumn{1}{c}{$\tau_{\rm TP-AGB}$} &
\multicolumn{1}{c}{$M_{\rm f}$} &
\multicolumn{1}{c}{$M_{\rm bol}^{\rm tip}$} &
\multicolumn{1}{c}{$M_{\rm bol}^{\rm peak}$} &
\multicolumn{1}{c}{$(L_{\rm EB}/L)_{\rm max}$} &
\multicolumn{1}{c}{$M_{\rm env}(\rm no\, EB)$} \\
\noalign{\smallskip}
\hline
\noalign{\smallskip}
0.019 & 1.68 & 3.5 & 0.683 & 1.623E+06 & 0.845 & -6.035 & 
\multicolumn{3}{c}{................No over-luminosity................} \\
      &      & 4.0 & 0.797 & 7.372E+05 & 0.896 & -6.211 & 
\multicolumn{3}{c}{................No over-luminosity................} \\
      &      & 4.5 & 0.877 & 4.312E+05 & 0.951 & -6.402 & 
\multicolumn{3}{c}{................No over-luminosity................} \\
      &      & 5.0 & 0.913 & 4.139E+05 & 0.996 & -6.558 & -6.525 &  0.010 &  2.991 \\
      & 2.00 & 3.5 & 0.683 & 2.056E+06 & 0.912 & -6.281 &
\multicolumn{3}{c}{................No over-luminosity................} \\
      &      & 4.0 & 0.797 & 1.099E+06 & 0.960 & -6.441 & -6.428 &  0.016 &  2.264 \\
      &      & 4.5 & 0.877 & 6.593E+05 & 0.998 & -6.568 & -6.604 &  0.065 &  1.978 \\
      &      & 5.0 & 0.913 & 5.129E+05 & 1.020 & -6.644 & -6.779 &  0.154 &  1.879 \\
      & 2.50 & 3.5 & 0.683 & 2.391E+06 & 0.979 & -6.518 & -6.570 &  0.067 &  1.465 \\
      &      & 4.0 & 0.797 & 1.253E+06 & 0.994 & -6.564 & -6.747 &  0.178 &  1.202 \\
      &      & 4.5 & 0.877 & 7.205E+05 & 1.013 & -6.645 & -6.909 &  0.265 &  1.119 \\
      &      & 5.0 & 0.913 & 5.427E+05 & 1.029 & -6.695 & -7.069 &  0.334 &  1.146 \\
\noalign{\smallskip}
\hline
\noalign{\smallskip}
0.008  &  1.68  & 3.5 & 0.756 & 1.451E+06 & 0.935 & -6.345 & 
\multicolumn{3}{c}{................No over-luminosity................} \\
 &        & 4.0 & 0.873 & 6.990E+05 & 0.987 & -6.521 & -6.500 &  0.011 &  2.239 \\
 &        & 4.5 & 0.918 & 5.327E+05 & 1.021 & -6.632 & -6.697 &  0.097 &  1.891 \\
 &        & 5.0 & 0.958 & 3.853E+05 & 1.043 & -6.720 & -6.859 &  0.175 &  1.905 \\
 &  2.00  & 3.5 & 0.756 & 1.711E+06 & 0.985 & -6.556 & -6.554 &  0.053 &  1.599 \\
 &        & 4.0 & 0.873 & 7.935E+05 & 1.008 & -6.618 & -6.745 &  0.164 &  1.268 \\
 &        & 4.5 & 0.918 & 5.645E+05 & 1.027 & -6.655 & -6.923 &  0.250 &  1.247 \\
 &        & 5.0 & 0.958 & 3.990E+05 & 1.047 & -6.721 & -7.083 &  0.316 &  1.234 \\
 &  2.50  & 3.5 & 0.756 & 1.779E+06 & 1.002 & -6.601 & -6.843 &  0.234 &  0.880 \\
 &        & 4.0 & 0.873 & 8.279E+05 & 1.019 & -6.662 & -7.018 &  0.328 &  0.751 \\
 &        & 4.5 & 0.918 & 6.025E+05 & 1.039 & -6.720 & -7.188 &  0.393 &  0.776 \\
 &        & 5.0 & 0.958 & 4.018E+05 & 1.054 & -6.745 & -7.343 &  0.450 &  0.662 \\
\noalign{\smallskip}
\hline
\noalign{\smallskip}
0.004 & 1.68 & 3.5 &  0.834 &  1.409E+06 &  0.986 &  -6.535 &  -6.535 &  0.000 &  1.721 \\
      &      & 4.0 &  0.893 &  9.694E+05 &  1.016 &  -6.627 &  -6.746 &  0.134 &  1.586 \\
      &      & 4.5 &  0.935 &  7.378E+05 &  1.041 &  -6.725 &  -6.929 &  0.220 &  1.447 \\
      & 2.00 & 3.5 &  0.834 &  1.517E+06 &  1.007 &  -6.613 &  -6.777 &  0.171 &  1.021 \\
      &      & 4.0 &  0.935 &  1.041E+06 &  1.031 &  -6.694 &  -6.970 &  0.270 &  1.109 \\
      &      & 4.5 &  0.893 &  7.736E+05 &  1.054 &  -6.769 &  -7.131 &  0.327 &  1.045 \\
      & 2.50 & 3.5 &  0.834 &  1.660E+06 &  1.045 &  -6.749 &  -7.082 &  0.305 &  0.639 \\
      &      & 4.0 &  0.893 &  1.165E+06 &  1.078 &  -6.862 &  -7.276 &  0.361 &  0.714 \\
      &      & 4.5 &  0.935 &  8.882E+05 &  1.102 &  -6.940 &  -7.463 &  0.427 &  0.551   \\   
\noalign{\smallskip}
\hline
\end{tabular}
\small
\begin{itemize}
\item $Z$: initial metallicity
\item $M_{\rm i}$: initial mass at the ZAMS ($M_{\odot}$)
\item $M_{\rm c,0}$: core mass at the onset of the TP-AGB phase ($M_{\odot}$)
\item $\tau_{\rm TP-AGB}$: TP-AGB lifetime (yr)
\item $M_{\rm f}$: final mass ($M_{\odot}$)
\item $M_{\rm bol}^{\rm tip}$: bolometric magnitude at the tip of the AGB
\item $M_{\rm bol}^{\rm peak}$: bolometric magnitude at the maximum efficiency
of envelope burning, before the onset of the super-wind phase
\item $(L_{\rm EB}/L)_{\rm max}$: maximum relative contribution of
envelope burning to the surface luminosity
\item $M_{\rm env}({\rm no \, EB})$: envelope mass at the extinction of envelope
burning ($M_{\odot}$)
\end{itemize}
\end{table*}

%

\subsection{The dependence on $Z$} 
\label{ebz}
At given $M$ and $\alpha$, the over-luminosity 
is more pronounced at lower metallicities.
It is worth remarking that the direct effect of chemical composition
on the base temperature is negligible itself (Sackmann \& Boothroyd 1991).
The greater efficiency of envelope burning is mostly
due to the fact that
a star of given $M$ enters the AGB phase with a core mass, $M_{\rm c}$,
that is greater for decreasing metallicity (see for example the
$M_{\rm c,0}$ entry in Table~\ref{hbbtab}). 
Considering that in the deepest
layers of the envelope close to the core,
 the radiative gradient 
depends like $\nabla_{\rm r} \propto M_{\rm c}^{\gamma}$,
where $\gamma$ is always positive ($0.5 \div 2.5$, as indicated
by Scalo et al. 1975),
it follows that  a deeper penetration of external convection
is expected for higher values of the core mass, and hence for lower metallicities.

\subsection{The dependence on $\alpha$} 
\label{eba}
\begin{figure*}
\centerline{
\psfig{file=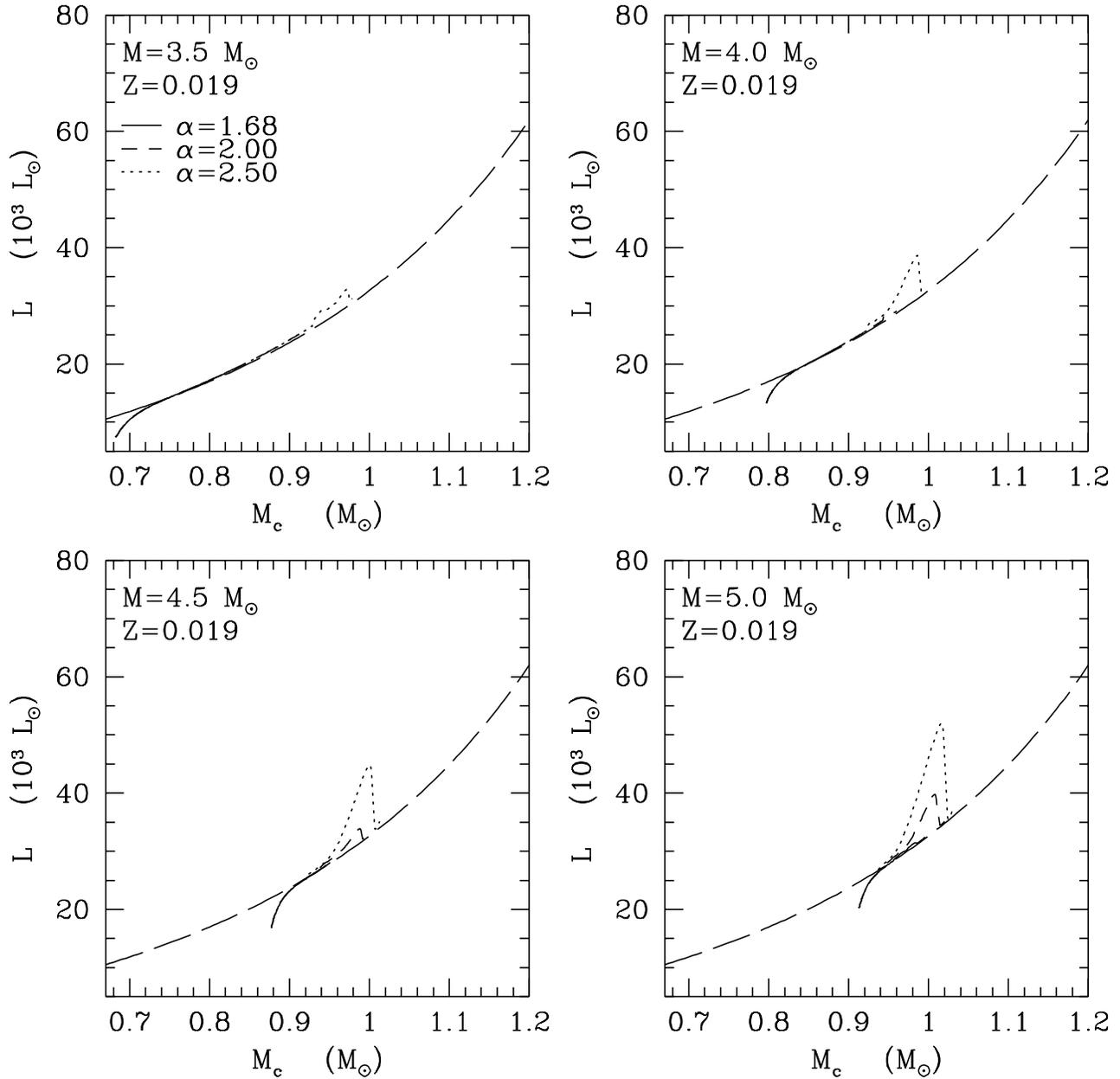,height=18truecm}}
\caption{Luminosity evolution of 3.5, 4.0, 4.5, 5.0 $M_{\odot}$ 
TP-AGB model stars 
with initial metallicity $Z=0.019$ as
a function of the core mass
 for three values of the mixing-length parameter, $\alpha$, as indicated. 
Stellar luminosities refer to the pre-flash maximum stage.
The long-dashed lines correspond to the reference standard core
mass-luminosity relation.}
\label{extralz02}
\end{figure*}

\begin{figure*}
\centerline{
\psfig{file=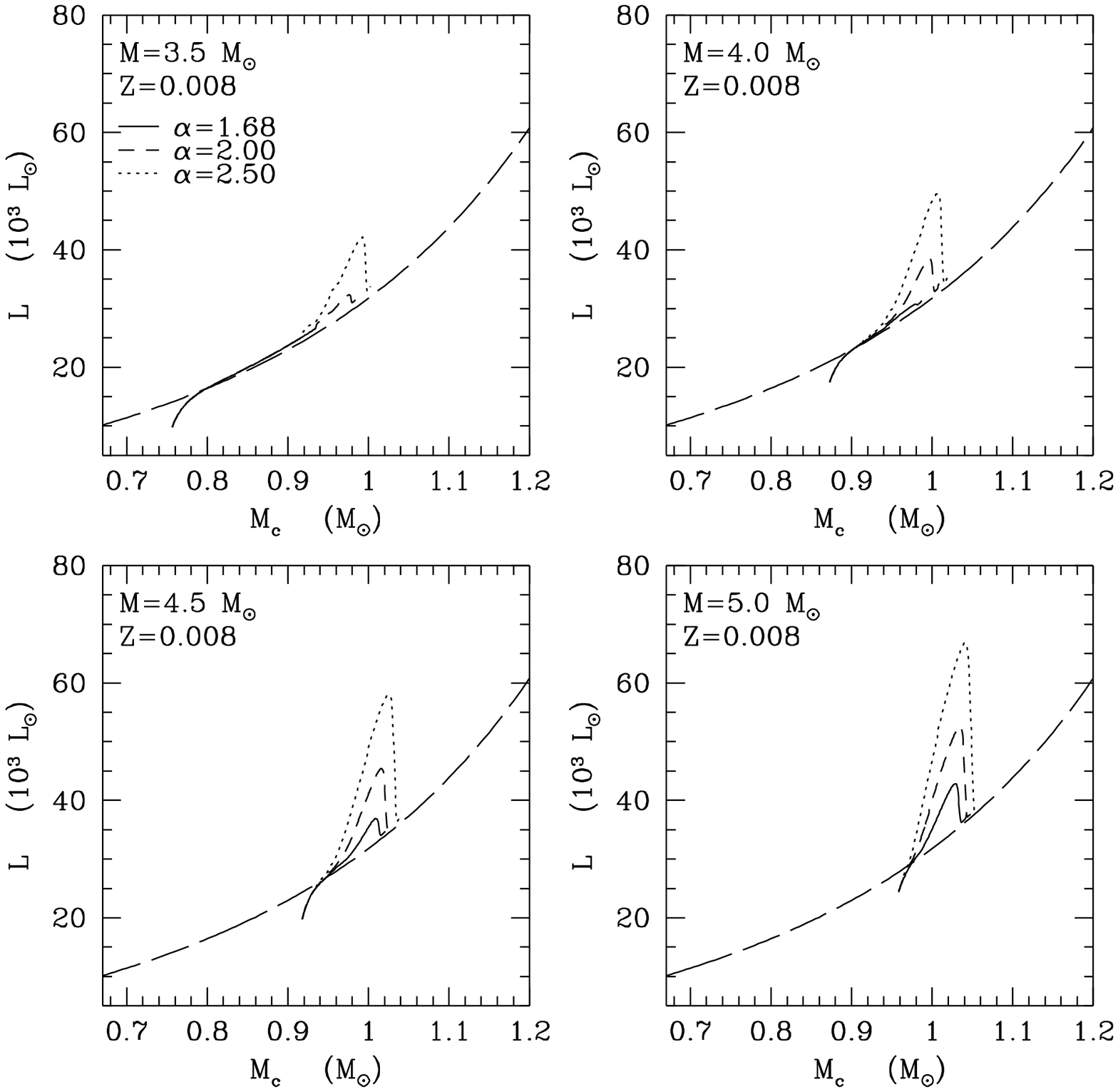,height=18truecm}}
\caption{ 
The same as in Fig.~\protect\ref{extralz02}, but with $Z=0.008$.
}
\label{extralz008}
\end{figure*}

\begin{figure*}
\centerline{
\psfig{file=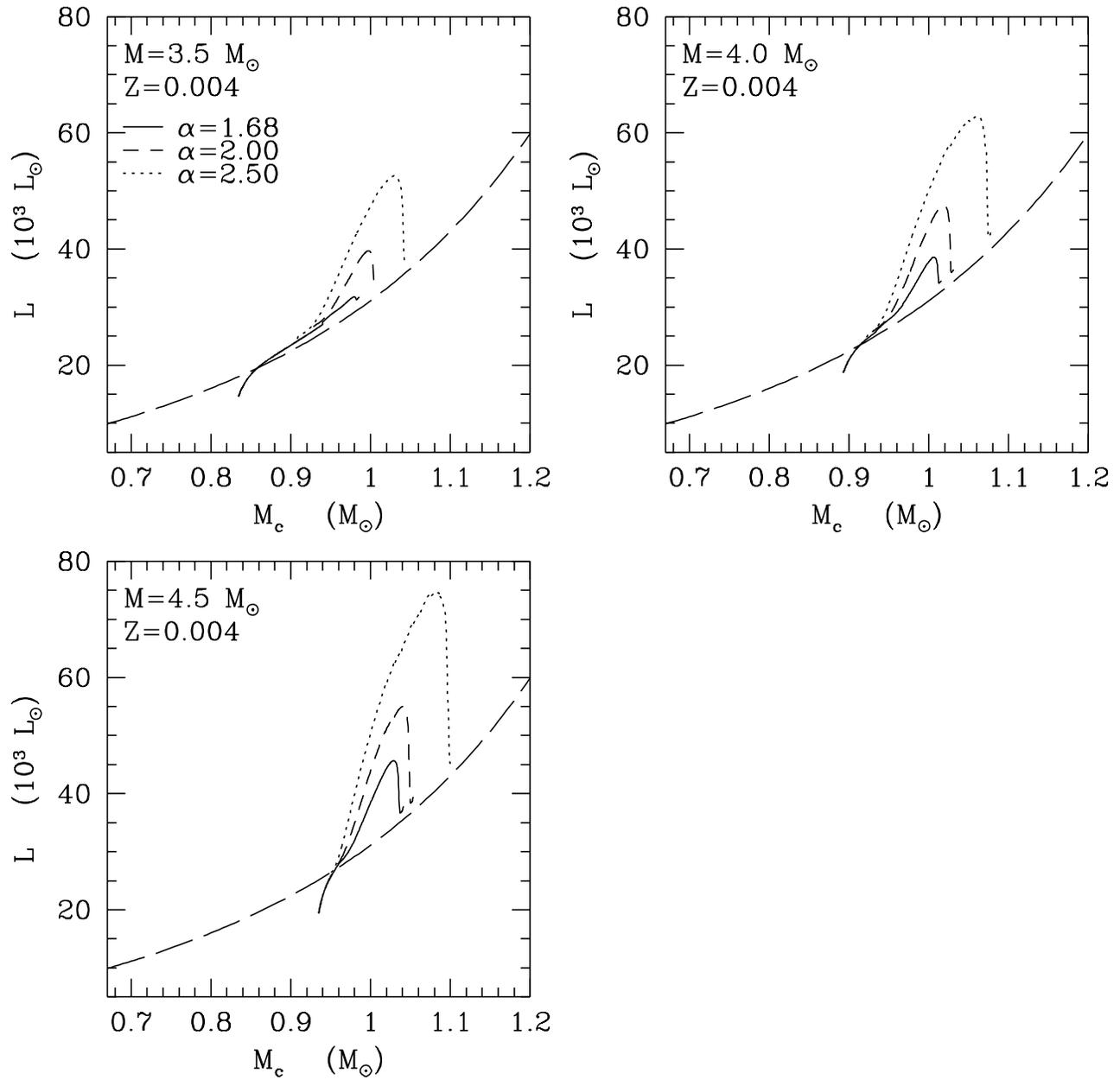,height=18truecm}}
\caption{ 
The same as in Fig.~\protect\ref{extralz02}, but with $Z=0.004$.
}
\label{extralz004}
\end{figure*}

The dependence 
of envelope burning on $\alpha$ is remarkable as already pointed
out by various authors (Sackmann \& Boothroyd 1991; D'Antona
\& Mazzitelli 1996). 
An increase of $\alpha$ 
causes an overall local rise
of the temperature profile across the envelope, so that
both the effective temperature, $T_{\rm eff}$, and the
base temperature, $T_{\rm b}$, are hotter.
It follows that $\alpha$ affects both the external configuration
 of an AGB star (i.e. the position on the H-R diagram),
and its internal structure (i.e. the temperature stratification  
of the convective envelope and related nuclear energy generation).
In particular, the higher temperatures, attained at the base 
of the convective envelope at increasing $\alpha$, determine a stronger
efficiency of nuclear burning. This is evident in
Figs.~\ref{extralz02} -- \ref{extralz004} from the 
greater amplitude of the luminosity excursion in models with the 
same mass and metallicity (see also the $M_{\rm bol}^{\rm tip}$ and
$(L_{\rm EB}/L)_{\rm max}$ entries in Table~\ref{hbbtab}).

However,  
changing the value of the mixing-length parameter 
 produces important consequences which are
not merely related to the efficiency of envelope
burning, but deal with further aspects of the evolution of these stars. 
 
A notable point concerns mass loss, and consequently the 
TP-AGB lifetimes (see the $\tau_{\rm TP-AGB}$ entry in Table~\ref{hbbtab}).
From our analysis it turns out that 
for a given model, the onset of the super-wind
regime is delayed for higher values of the mixing-length parameter.
Consequently, for given initial mass and metallicity 
the whole duration of the TP-AGB phase
is longer for a star with a more efficient envelope burning 
obtained by increasing $\alpha$.
This result does not contradict the result that 
the high luminosities 
produced by envelope burning would anticipate the end 
the evolution by triggering enhanced mass-loss rates 
(see for example BS91). 
To clarify this point, let us consider the competition between
luminosity and effective temperature in determining the 
efficiency of mass loss.
According to Vassiliadis \& Wood's prescription  (1993),
adopted in  our calculations, the dependence of the mass-loss rate 
before the development of the superwind, can be expressed
as $\log\dot M \propto P \propto T_{\rm eff}^{-3.88} L^{0.97} M^{-0.9}$,
where $P$ is the fundamental period of pulsation.
Hence,  an increase of $\alpha$  operates  
in two opposite directions.
From one side, it favours mass loss via the term $L^{0.97}$,
as it
strengthens the efficiency of envelope burning, 
quickly leading to higher luminosities.
From the other side, it weakens mass loss via the term
$T_{\rm eff}^{-3.88}$, 
as it reduces the stellar radius, yielding higher
values of the effective temperature.
The net result depends on the prevailing effect, that in our case
turns out to be related to the increase of the effective temperature. 
It is interesting to notice, however, that different formulations for mass loss
could produce different results, owing to their specific dependence
on stellar parameters, i.e. luminosity and effective temperature.

\clearpage
%
%

%

\clearpage
\section{Concluding remarks}
\label{end}

%
In this study it has been shown that 
the break-down of the $M_{\rm c}-L$ relation and the related   
over-luminosity effect produced by the occurrence of envelope burning
in the most massive TP-AGB stars can be consistently taken into account
in synthetic calculations. In fact, the original method based on
envelope integrations is able to successfully reproduce the
results of complete evolutionary models.

This nice result  
 strengthens   
the reliability of envelope integrations in synthetic 
TP-AGB models for various related aspects, such as
the nucleosynthesis occurring 
in hot bottom envelopes and the chemical yields.

The results of the present paper derive from extensive calculations
of the AGB phase carried out for a fine grid of stellar masses
($0.8 M_{\odot} \la M \le 5 M_{\odot}$) and various metallicities
($Z=0.019,\, Z=0.008$, and $Z=0.004$).
This work is actually a part of a project
aimed at investigating various
aspects of the AGB evolution by means of   
a flexible and accurate synthetic model.
If the flexibility is an intrinsic characteristic of synthetic codes,
this model gains in accuracy  thanks to the use of 
analytical prescriptions derived from detailed 
AGB calculations (e.g. Wagenhuber 1996), coupled  
to a complete envelope model updated with recent input physics
(i.e. opacities, nuclear reactions rates).

Up-coming papers are devoted to present and analyse other issues of interest.
In particular, basing on  
the improved treatment of the third dredge-up included in
synthetic calculations (briefly outlined in Sect.~\ref{dred3}) we will
address the question of reproducing the luminosity functions of carbon stars 
in both the LMC and SMC (Marigo et al. 1998, in preparation).
A  further work will be dedicated to present the results on the   
predicted changes in the surface chemical abundances
of AGB stars and related chemical yields, as a function of the
stellar mass and metallicity (Marigo et al. 1998, in preparation).     
  

\begin{acknowledgements}
The author is grateful to Alessandro\ Bressan, Cesare\ Chiosi, and 
L\'eo Girardi for their important advice and 
kind interest in this work. Many thanks to Achim Weiss for carefully
reading the manuscript and for his useful remarks.  
Thomas Bl\"ocker is acknowledged for providing the
results of his calculations. 
\end{acknowledgements}

\end{document}